# High Voltage Ride Through Strategy for DFIG Considering HVDC System Converter Blocking

Changping Zhou, Zhen Wang, Ping Ju, and Deqiang Gan

*Abstract*—This paper presents a *P-Q* coordination based high voltage ride through (HVRT) control strategy for double fed induction generators (DFIGs) based on a combined *Q-V* control and *P-V* de-loading control. The active/reactive power injection effect of DFIG on transient overvoltage is firstly analyzed and the reactive power capacity evaluation of DFIG considering its de-loading operation is then conducted. In the proposed strategy, the reactive power limit of DFIG can be flexibly extended during the transient process in coordination with its active power adjustment, as a result the transient overvoltage caused by DC bipolar block can be effectively suppressed. Moreover, key outer loop parameters such as *Q-V* control coefficient and de-loading coefficient can be determined based on the point of common coupling (PCC) voltage level and the available DFIG power capacity. Finally, case studies based on MATLAB/Simulink simulation are used to verify the effectiveness of the proposed control strategy.

*Index Terms*—Active power de-loading control, high voltage ride through (HVRT), reactive power absorption, transient overvoltage.

## I. INTRODUCTION

THE past ten years have witnessed the rapid development of wind power for the purpose of solving the worldwide traditional energy and environmental crises [1]. Due to the inherent uncertainty and volatility of wind power resources, the necessary power electronic interfaces for integrating wind turbine generators (WTGs) has posed big challenges to power system operation and control, such as additional reserve requirement and inertial reduction in weak connection [2].

In Northwest China, many large-scale wind farms are located far away from the receiving-end grids, and thus forming a weak-grid integration [3]. For de facto long-distance transmission of wind power in China, high voltage direct current (HVDC) transmission technology is widely used for grid connection. Since each HVDC converter can consume a large amount of reactive power (about 40%-60% of the transferred active power) and additional reactive power compensators (RPCs) such as a group of capacitor banks are needed to provide enough reactive power required for commutation. When any DC bipolar block occurs, surplus reactive power due to delayed compensators removing, will cause transient overvoltage surge [4]. In this case, the WTG overvoltage protection will evoke further WTG shedding, and even system instability risk [5]. In [6], an overvoltage phenomena is observed in offshore wind farms following blocking of the connected HVDC converter. The mechanism of transient overvoltage caused by DC blocking and its influence factors are analyzed in [7] and further a short circuit ratio (SCR) based index is proposed to measure the transient overvoltage level.

Fortunately, many efforts have been devoted to relieve above transient overvoltage problem on wind farm side, which can be approximately classified into three categories. The first one can be attributed into a type of external RPC control methods, in which additional RPCs or their combination on the stator side are equipped and RPC control schemes are designed accordingly, e.g., a combination scheme of static synchronous compensator and dynamic voltage restorer is proposed to achieve DFIG fault ride through in [8], [9]; a superconducting fault current limiter and a transient voltage controller are jointly used to improve the system transient voltage stability in [10], [11]. The second one is a type of rotor-side converter (RSC) supplementary circuit control methods, where protection circuits are added in the rotor or DC circuit to attenuate the rotor overcurrent or DC overvoltage, e.g., a series dynamic braking resistor based crowbar in rotor circuit to limit current surge in [12]; a DC series chopper circuit to maintain a constant DC voltage in [13]; a DC series storage protection circuit to store excess energy during voltage surge in [14]. The third one is a type of RSC/GSC (grid side converter) control methods, e.g., a RSC demagnetizing current controller to suppress inner voltage saturation is developed in [15], [16]; a GSC high voltage ride through (HVRT) control strategy with adaptive adjustment of DC-link voltage reference is proposed in [17]; an automatic voltage control at PCC based on optimized RSC reactive current reference to maximize the reactive power support capability of DFIG is applied in [18], [19].

However, the reactive power capacity of DFIG has not been fully explored in above solutions with active power regulation taken into account. In this paper, the effect of the active/reactive power injection of WTG on overvoltage is firstly studied and the reactive power limit of DFIG considering its de-loading operation is analyzed subsequently. Then, a *P-Q* coordination method is proposed for emergency reactive power control to suppress transient overvoltage, in which a *Q-V* control logic is developed for reactive power adjustment and a *P-V* de-loading control logic is presented for active power adjustment accordingly. As a result, the reactive power supporting potential

Manuscript received: May 12, 2019; accepted: October 23, 2019. This work was jointly supported by the National Nature Science Foundation of China (No. 51677165, No. 51837004) and the National Key R&D Program of China (No. 2017YFB0902000).

The authors are with the College of Electrical Engineering, Zhejiang University, Hangzhou 310027, China (e-mail: 11610027@zju.edu.cn,eezwang@ieee.org, dgan@zju.edu.cn, pju@hhu.edu.cn). Corresponding author: Zhen Wang can be enlarged flexibly. The main contribution of this paper is summarized as follows.

1) The reactive power operation region of DFIG considering its active power de-loading control is elaborately depicted.

2) A *P-Q* coordination based HVRT strategy is proposed to fulfill transient overvoltage suppression, in which the *Q-V* control is coordinated with the *P-V* de-loading control to achieve the maximum reactive power capacity.

The rest of this paper is organized as follows: Section II analyzes the sending-end transient overvoltage and the effect of the active/reactive power injection of DFIG on transient overvoltage as well as the reactive power limit evaluation of DFIG. Section III introduces the proposed *P-Q* coordination based HVRT control strategy and the determination of key parameters. Section IV provides the case studies. Finally, conclusions are drawn in Section V.

## II. TRANSIENT OVERVOLTAGE ANALYSIS

### A. System Description

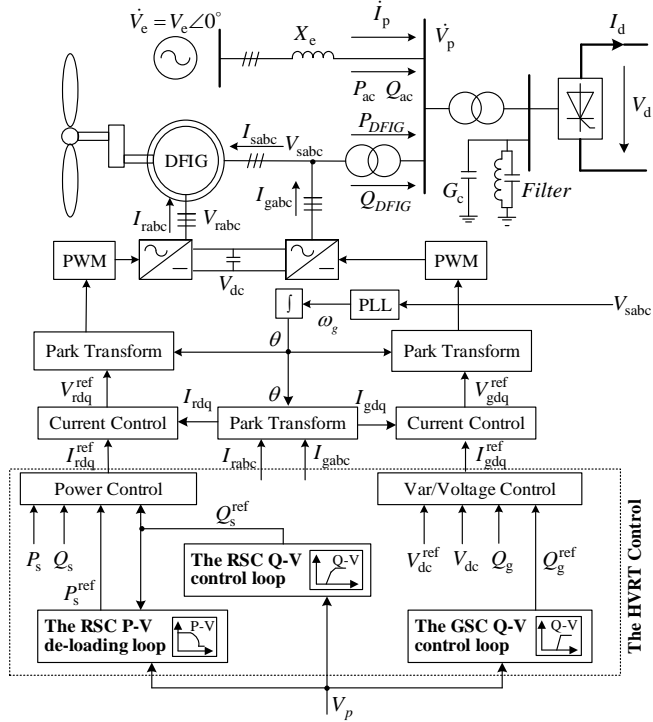

Fig. 1 Illustration of a sending-end AC system and HVRT control

The DFIG-based wind farm integrated system in Fig. 1 represents a typical case of bulk wind power transmission from generation centers to load centers by means of HVDC transmission. An equivalent model consisting of the equivalent reactance $X_e$ and voltage source $\dot{V}_e$ is used to represent the sending-end AC grid. For simplicity purposes, a single DFIG-based wind turbine is used to represent the aggregate behavior of the whole wind farm. The traditional double loop control structure is applied for DFIG converter control. Originally under normal operating condition, the conventional maximum power point tracking (MPPT) control is used for the RSC outer loop and the constant *Q-V* control is used for the GSC outer loop to maintain the DC-link voltage. The outer loop will generate the reference signals for the inner current loop independently. The inner current control is carried out in rotational framework (*d-q*) with stator flux orientation with the help of the phase-locked-loop (PLL). Basic proportional-integral (PI) controller is adopted in those inner control loops [20], [21]. Moreover, a group of static capacitor and filter is installed at the rectifier station to compensate for the reactive power consumption.

Under the overvoltage operation condition with the DC bipolar blocking described below, the following outer loops are designed and applied: ① two *Q-V* independent control loops are designed for the RSC power control and the GSC *Q-V* control, respectively [22], [23]; ② a *P-V* de-loading loop is designed to generate the power reference for active power control. The details will be elaborated in next section.

### B. Effect of Active/Reactive Power Injection of WTG

When a transient overvoltage surge happens at the sending-end AC system in Fig. 1, the terminal voltage at PCC $V_p$ satisfies the following relationship:

$$aV_p^4 + bV_p^2 + c = 0 \quad (1)$$

where $a = (1 - G_c X_e)^2$, $G_c$ is the admittance of HVDC capacitor banks; $b = -1 - 2(1 - G_c X_e)Q_{DFIG}X_e$, $Q_{DFIG}$ is the total reactive power of DFIG; and $c = (P_{DFIG}^2 + Q_{DFIG}^2)X_e^2$, $P_{DFIG}$ is the total active power of DFIG. The derivation of (1) can be found in Appendix A.

Accordingly, the sensitivity of $V_p$ with respect to $P_{DFIG}$ and $Q_{DFIG}$ can be formulated from (1) as:

$$\frac{\partial V_p}{\partial P_{DFIG}} = -\frac{1}{V_p} \frac{P_{DFIG}X_e^2}{\sqrt{b^2 - 4ac}} \quad (2)$$

$$\frac{\partial V_p}{\partial Q_{DFIG}} = \frac{(1 - G_c X_e)X_e(1 + \sqrt{b^2 - 4ac})}{2aV_p\sqrt{b^2 - 4ac}} \quad (3)$$

It is obvious from (2) that $\partial V_p / \partial P_{DFIG} < 0$ when $P_{DFIG} > 0$.

If the sending-end AC system has SCR greater than 0.6, then the following conclusions can be drawn from Appendix A:

1) $\partial V_p / \partial Q_{DFIG} > 0$. In other words, when overvoltage at PCC happens, the absorbing reactive power of wind farm (in inductive operation) can help suppress the transient overvoltage.

2) $\left| \partial V_p / \partial Q_{DFIG} \right| > \left| \partial V_p / \partial P_{DFIG} \right|$, which means adjusting the reactive power absorption of DFIG has the better overvoltage suppression effect than adjusting the active power.

### C. Inductive Reactive Power Limit

*1) Reactive Power Evaluation*

The inductive reactive power of wind farm comes from two resources: DFIG stator by RSC control and GSC itself.

The stator reactive power limit is mainly determined by: ① the maximum rotor current constraint in (4) [24]; ② the generator capacity constraint in (5) [25], [26]. The details can be found in Appendix B.

$$P_s^2 + (Q_s + \frac{3V_s^2}{2X_l})^2 \leq (\frac{3X_m}{2X_l}V_s I_r^{max})^2 \quad (4)$$

$$P_s^2 + Q_s^2 \leq S_n^2 \quad (5)$$

where $P_s$ and $Q_s$ are the active power and reactive power from DFIG stator; $V_s$ is the DFIG stator voltage; $X_l = X_{ls} + X_m$ is the stator inductance, $X_{ls}$ and $X_m$ are the DFIG stator leakage and magnetizing reactance; $I_r^{max}$ is the maximum current limit of rotor; and $S_n$ is the apparent power capacity of DFIG.

Therefore, the inductive reactive power limit of stator $Q_s^{max}$ can be determined by the less value between the reactive power corresponding to the maximum rotor current limit $Q_{s_1}^{max}$ and the maximum available reactive power $Q_{s_2}^{max}$:

$$\begin{cases} Q_{s_1}^{max} = 1.5V_s^2/X_l + \sqrt{(1.5X_m V_s I_r^{max}/X_l)^2 - P_s^2} \\ Q_{s_2}^{max} = \sqrt{S_n^2 - P_s^2} \\ Q_s^{max} = \min\{Q_{s_1}^{max}, Q_{s_2}^{max}\} \end{cases} \quad (6)$$

An illustration of the relationship between $Q_s^{max}$ and its active power $P_s$ is given in Fig. 2(a). It is clear when $P_s$ is less than $P_0$, $Q_s^{max} = Q_{s_1}^{max}$; otherwise $Q_s^{max} = Q_{s_2}^{max}$.

The characteristic diagram of the stator inductive reactive power operating region with respect to $P_s$ and $V_s$ at the rated wind speed is plotted in Fig. 2(b) and it can be observed that the immediate rise in stator voltage and the decrease of DFIG active power can both increase its reactive power limit.

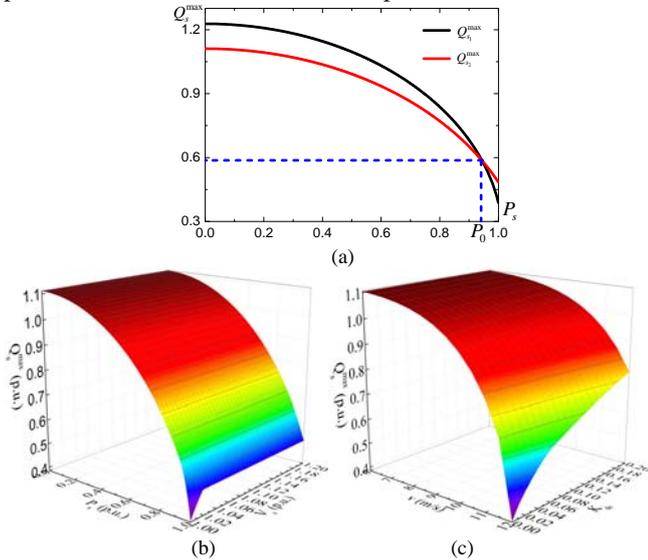

Fig. 2 Diagram of reactive power operation region of DFIG stator. (a) Reactive power limit curve of stator as its active power changes. (b) $Q_s^{max}(P_s, V_s)$. (c) $Q_s^{max}(v, K_{de})$.

On the other hand, the reactive power limit of GSC is determined by its apparent power capacity:

$$Q_g^{max} = \sqrt{S_c^2 - (sP_s)^2} \quad (7)$$

where $S_c$ is GSC apparent power capacity, $s$ is the speed slip.

In general, GSC is controlled to maintain the DC voltage and the its active power output $sP_s$ can be neglected. As a result, $Q_g^{max}$ can be approximated to be $S_c$ [18], [27] and the total inductive reactive power limit $Q_G^{max}$ that a DFIG can absorb is determined by:

$$Q_G^{max} = Q_s^{max} + Q_g^{max} \quad (8)$$

*2) Active Power De-loading*

From (6), it is clear that the reactive power limit can be extended by the decrease of DFIG active power, i.e., DFIG de-loading operation. The principle of DFIG de-loading is given in Fig. 3, in which initially DFIG is at MPPT operation point $a$, and then it actively traces some de-loading curve defined in (9) to point $b$ so that part of available wind energy can be stored in form of rotating kinetic energy [28], [29]. In Fig. 3, the corresponding active power after de-loading is as follows.

$$P_s = P_{de} = (1 - K_{de})P_{MPPT} \quad 0 \leq K_{de} < K_{de}^{max} \quad (9)$$

where $P_{MPPT}$ and $P_{de}$ are DFIG active power under MPPT and de-loading modes, respectively; $K_{de}$ is de-loading coefficient and $K_{de}^{max}$ is its maximum value that limits the minimum active power output of DFIG under de-loading operation to avoid the rotor overspeed. As far as DFIG is concerned, $K_{de}^{max}$ is about 0.1-0.2[29], [30] and $K_{de}^{max} = 0.2$ is set in the study.

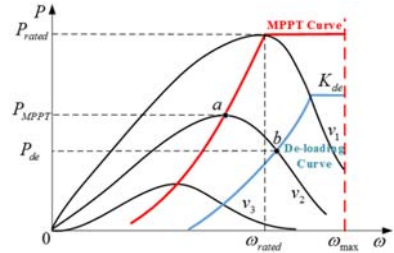

Fig. 3 Illustration of DFIG de-loading operation

Further, the inductive reactive power operation region of DFIG with different wind speeds $v$ and de-loading coefficients is plotted in Fig. 2(c). It can be observed: ① normally the reactive power limit will be reduced with the increase of wind speed; ② the limit can be enlarged by active power de-loading control, especially in high wind speed range (10-12 m/s). Thus, the power limits of the stator and DFIG under de-loading $Q_{sD}^{max}$, $Q_{GD}^{max}$ can respectively be updated from (6)-(9) accordingly.

III. A *P-Q* COORDINATION BASED HVRT CONTROL

Since any DFIG converter capacity is limited, the main idea to capture additional reactive power capacity is by means of reactive power control and active power de-loading control coordinately, which will be elaborated below.

*A. P-Q Coordination*

The *P-Q* coordination includes two main logics: ① stator and GSC reactive power absorption by RSC and GSC reactive power control; ② reactive power capacity enlargement by de-loading control. The main procedure of the *P-Q* coordination based HVRT control is illustrated in Fig. 4. When the overvoltage condition is detected, firstly the reactive power limit is evaluated and the outer loop parameters are determined according to the rules explained below; secondly, RSC and GSC reactive power control is executed to absorb the surplus reactive power; thirdly, the DFIG de-loading will be executed and the resultant reactive power capacity will be enlarged. The whole procedure will continue until the PCC voltage is satisfied or the maximum de-loading is reached.

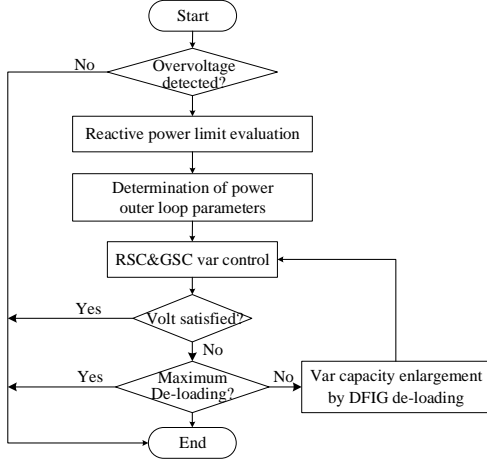

Fig. 4 Flow chart of P-Q coordination

*1) Q-V Control Logic*

For the two reactive control logics, a *Q-V* piecewise control rule is adopted and illustrated in Fig. 5(a).

$$|Q_{DFIG}| = \begin{cases} 0 & V_p \leq V_{OV}^{\min} \\ K_1(V_p - V_{OV}^{\min}) & V_{OV}^{\min} < V_p \leq V_{OV1} \\ Q_G^{\max} + K_2(V_p - V_{OV1}) & V_{OV1} < V_p \leq V_{OV}^{\max} \\ Q_{GD}^{\max} & V_p > V_{OV}^{\max} \end{cases} \quad (10)$$

where $V_{OV}^{\min}$ and $V_{OV}^{\max}$ are the minimum and maximum voltages to activate HVRT; $V_{OV1}$ is the de-loading triggering voltage in the HVRT curve; and $Q_{DFIG}$ <0 means DFIG will absorb reactive power.

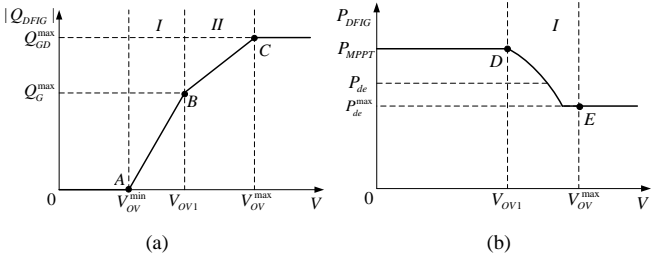

(a)                                (b)

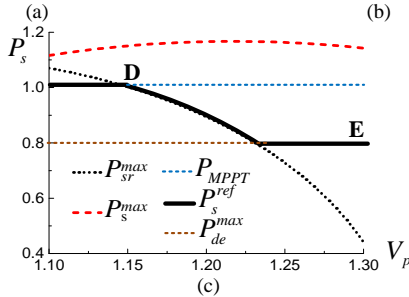

(c)

Fig. 5  *P-Q* coordination HVRT control. (a) *Q-V* piecewise control curve. (b) *P-V* power control curve. (c) A snapshot of de-loading *P-V* trajectory.

As recommended by the test procedure of wind turbine HVRT capability in China [31], $V_{OV}^{\min}$ =1.1 and $V_{OV}^{\max}$ =1.3 are set in the paper. Initially $V_{OV1}$ =1.15 and its effect on the transient overvoltage will be discussed in next section.

As of *Q-V* piecewise control in Fig. 5(a), the whole process corresponds to two stages:

1) Stage 1: $V_{OV}^{\min} < V_p \leq V_{OV1}$, the RSC and GSC reactive power control is activated and the stator and GSC will absorb reactive power proportional to their reactive power limits along line *AB*. This process continues until the reactive power limit of DFIG is reached. The following reference reactive power setting is used:

$$\begin{cases} |Q_s^{ref}| = \dfrac{Q_s^{\max}}{Q_G^{\max}} K_1(V_p - V_{OV}^{\min}) & Q_s^{ref} < 0 \\ |Q_g^{ref}| = \dfrac{Q_g^{\max}}{Q_G^{\max}} K_1(V_p - V_{OV}^{\min}) & Q_g^{ref} < 0 \end{cases} \quad (11)$$

where $Q_s^{ref}$ and $Q_g^{ref}$ are the reference reactive powers for DFIG stator and GSC.

2) Stage 2: $V_{OV1} < V_p \leq V_{OV}^{\max}$, the reactive power limit of DFIG will be temporarily enlarged by active power de-loading control. This process continues until the maximum de-loading is reached. The reference power is set as follows:

$$\begin{cases} |Q_s^{ref}| = Q_s^{\max} + K_2(V_p - V_{OV1}), & Q_s^{ref} < 0 \\ |Q_g^{ref}| = Q_g^{\max}, & Q_g^{ref} < 0 \end{cases} \quad (12)$$

*2) P-V De-Loading Control Logic*

For the de-loading control logic in Fig. 5(b), where $P_{de}^{\max}$ is the active power of DFIG under de-loading mode, a *P-V* piecewise control rule is used:

$$P_{DFIG} = \begin{cases} P_{MPPT} & V_p \leq V_{OV1} \\ P_{de} & V_{OV1} < V_p \leq V_{OV}^{\max} \\ (1 - K_{de}^{\max}) P_{MPPT} & V_p > V_{OV}^{\max} \end{cases} \quad (13)$$

Normally, DFIG will operate at MPPT state when $V_p \leq V_{OV1}$. Otherwise, the de-loading control will be activated and DFIG will trace the de-loading curve *DE* until the maximum limit is reached.

The power outer loops corresponding to above mentioned *Q-V* control and *P-V* de-loading control logics are illustrated in Fig. 6, in which the outer loop parameters are discussed below.

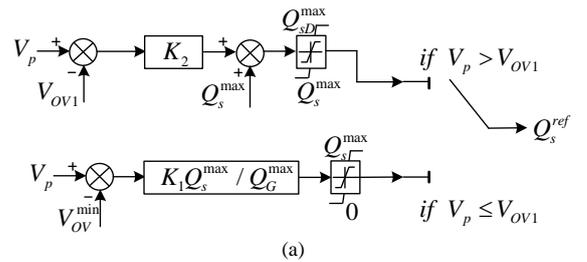

(a)

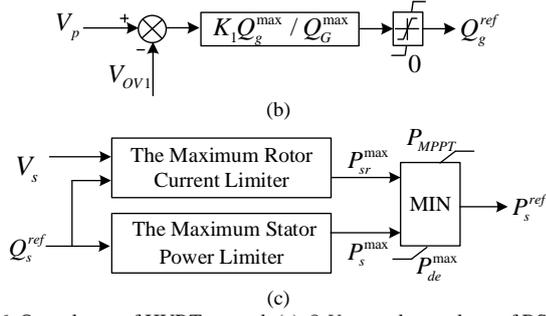

(b)

(c)

Fig. 6 Outer loops of HVRT control. (a) *Q-V* control outer loop of RSC. (b) *Q-V* control outer loop of GSC. (c) *P-V* de-loading outer loop.

### B. Determination of Outer Loop Parameters

*1) Q-V Control Coefficients $K_1$ and $K_2$*

As the control coefficients represent slopes of *Q-V* piecewise curve in Fig. 5(a), they can be easily obtained as:

$$\begin{cases} K_1 = Q_G^{\max} / (V_{OV1} - V_{OV}^{\min}) \\ K_2 = (Q_{GD}^{\max} - Q_G^{\max}) / (V_{OV}^{\max} - V_{OV1}) \end{cases} \quad (14)$$

*2) De-loading Coefficient $K_{de}$*

The de-loading logic in the power outer loop is illustrated in Fig. 6(c). Once $Q_s^{ref}$ is given in (11) and (13), the reference active power of de-loading $P_s^{ref}$ can be determined by:

$$P_s^{ref} = \min\{P_{sr}^{\max}, P_s^{\max}\} \quad (15)$$

$$P_{sr}^{\max} = \sqrt{(1.5X_m V_s I_r^{\max} / X_l)^2 - (Q_s^{ref} - 1.5V_s^2 / X_l)^2} \quad (16)$$

$$P_s^{\max} = \sqrt{S_n^2 - (Q_s^{ref})^2} \quad (17)$$

where $P_{sr}^{\max}$ is the active power corresponding to the maximum rotor current limit; and $P_s^{\max}$ is the maximum stator power limit.

Accordingly, the de-loading coefficient $K_{de}$ is:

$$K_{de} = 1 - P_{de} / P_{MPPT} \quad (18)$$

An example of the de-loading *P-V* trajectory $P_s^{ref}$ in Fig. 5(b) is plotted in Fig. 5(c) with the typical parameters in Table 1-Table 3. It can be found in Fig. 5(c) that the $P_s^{\max}$ will dominate during the whole overvoltage process.

Table 1 Test system parameters

| Parameter | Symbol | Value |
|---|---|---|
| Rated stator voltage | $V_{sN}$ | 690 V |
| Base frequency | $f_B$ | 50 Hz |
| Rated DFIG active power | $P_n$ | 150MW |
| Rated DFIG power factor | $\cos\varphi$ | 0.9 |
| Apparent power capacity of GSC | $S_c$ | 50 MVA |
| Mutual inductance | $X_m$ | 2.06 Ω |
| Stator inductance | $X_l$ | 2.24 Ω |
| Maximum current limit of RSC | $I_r^{\max}$ | 1600 A |
| Rated wind speed | $v_{rate}$ | 12 m/s |
| Rated voltage of HVDC | $V_{DC}$ | 500 kV |
| Rated DC current of HVDC | $I_{DC}$ | 2 kA |
| Reactive power compensation of rectifier station | $Q_c$ | 1000 Mvar |

Table 2 Parameters of DFIG PI controllers

| PI controller | $K_p$ | $K_i$ |
|---|---|---|
| RSC active power control | 2 | 20 |
| RSC reactive power control | 1 | 20 |
| RSC current control | 0.6 | 100 |
| GSC DC voltage control | 8 | 400 |
| GSC reactive power control | 2 | 20 |
| GSC current control | 0.83 | 100 |

Table 3 Control parameters of DFIG on initial condition

| Parameter | Value (p.u.) |
|---|---|
| $Q_s^{\max}$ | 0.48 |
| $Q_G^{\max}$ | 0.73 |
| $V_{OV1}$ | 1.15 |
| $Q_g^{\max}$ | 0.25 |
| $Q_{GD}^{\max}$ | 1.02 |
| $K_1$ | 9.6 |
| $K_2$ | 3.6 |

## IV. CASE STUDIES

### A. System Description

Figure 7 shows the studied system of DFIG-based wind farm integrated into AC Grid 2 by means of HVDC transmission. DFIG operates at the unit power factor under rated wind speed 12 m/s initially and the block occurs at $t = 0.5$ s. Parameters of the system and DFIG PI controllers are given in Table 1 and 2. Simulations are conducted in MATLAB/Simulink.

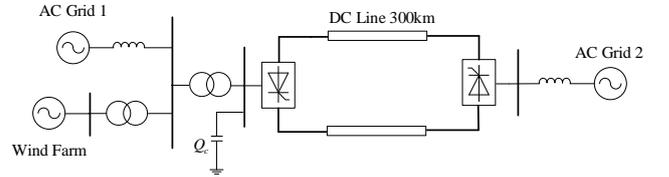

Fig. 7 Diagram of HVDC system with wind farms

With system parameters available in Table 1, the control parameters of DFIG on the initial steady state condition where $V_p = 1.0$ can be obtained as listed in Table 3.

### B. Method Comparison Study

In this section, a comparison study is conducted among the following four methods: ① method 1, the unit power factor control without reactive control; ② method 2, the automatic voltage control at PCC [19]; ③ method 3, the demagnetizing current control [16]; ④ method 4, the proposed *P-Q* coordination control. The resultant curves include the transient voltages at PCC and DC-link capacitor, the RSC voltage and current, the DFIG active power and reactive power are illustrated in Fig. 8.

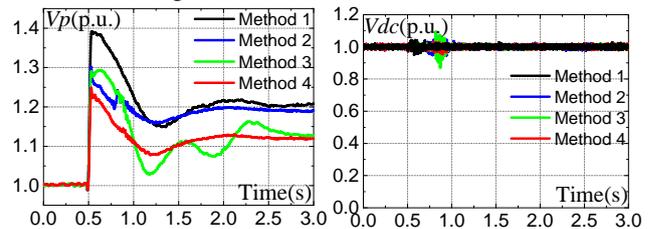

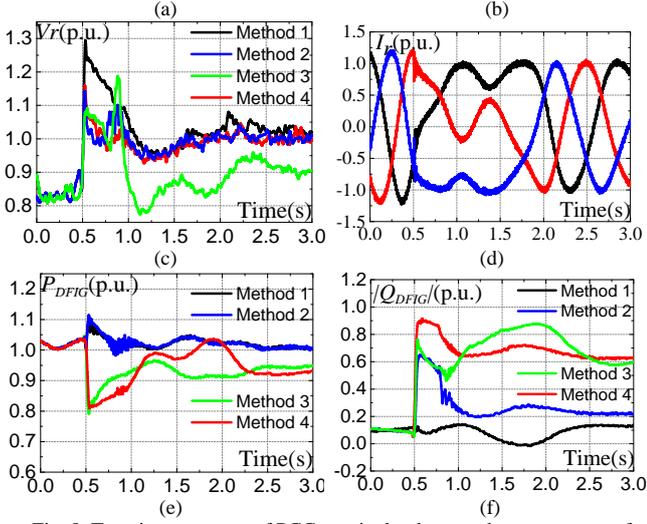

Fig. 8 Transient response of PCC terminal voltage and power output of DFIG. (a) Transient voltage at PCC. (b) Voltage of DC-link capacitor. (c) RSC voltage. (d) RSC current with method 4. (e) Active power of DFIG. (f) Reactive power of DFIG.

It can be observed from Fig. 8 that: ① Both RSC voltage/current and voltage of DC-link capacitor are within their limits under methods 2, 3, 4; ② Methods 2, 3, 4 can better reduce the overvoltage peak value compared with method 1, which can make the PCC voltage being below 1.3 p.u. to avoid further islanding risk; ③ method 4 has the best performance on overvoltage suppression compared with methods 2 and 3, e.g., the former has less peak/steady-state voltage and faster voltage recovery at PCC, because more reactive power can be absorbed with method 4. In summary, method 4 can effectively suppress transient overvoltage to help DFIG ride through the high voltage process after DC blocking and has the best performance compared with method 2 and 3.

### C. Parameter Effect Study

Firstly, the effect of wind speed on transient overvoltage with method 4 is studied. Simulations with three typical wind speeds (8, 10 and 12 m/s) are compared in Fig. 9(a). It shows that method 4 has better overvoltage suppression effect at lower speed because DFIG has larger reactive power capacity and thus can absorb more reactive power.

Further, the effect of de-loading triggering voltage $V_{OV1}$ on transient overvoltage with method 4 is investigated and the result is given in Fig. 9(b). It is found that less $V_{OV1}$ can achieve less steady-state voltage at PCC, because less $V_{OV1}$ means earlier de-loading control will be taken and more reactive power will be absorbed by DFIG.

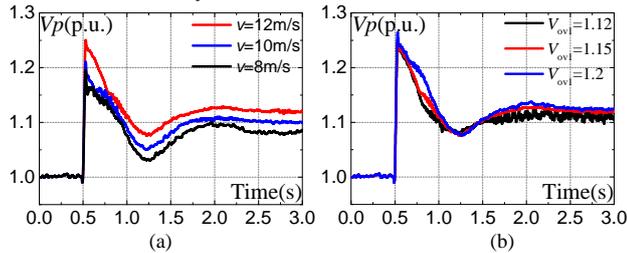

Fig. 9 Transient PCC voltage with different parameters. (a) Different wind speeds. (b) Different de-loading triggering voltages.

## V. CONCLUSION

This paper studies the effect of the active/reactive power injection of WTG on transient overvoltage and proposes a *P-Q* coordination based HVRT control strategy for DFIG by means of a combined *Q-V* control and de-loading control to suppress PCC overvoltage. In the strategy, the reactive power operation region of DFIG considering its active power de-loading control is firstly evaluated and the outer loop parameters are determined according to the rules explained. Then, the RSC and GSC reactive power control is executed to absorb the surplus reactive power. The reactive power capacity of DFIG can be flexibly enlarged to absorb more reactive power in coordination with rapid active power control of DFIG to achieve de-loading operation. Finally, the simulations verify that the proposed HVRT strategy outperforms on PCC overvoltage suppression to avoid DFIG islanding compared with other two control methods.

## APPENDIX A

### A. Sensitivity Analysis

An illustration of the voltage relationship between $\dot{V}_p$ and $\dot{V}_e$ is presented in Fig. A1(a) and is obtained as:

$$\begin{cases} \dot{V}_p = \dot{V}_e - \Delta \dot{V} = \dot{V}_e - jX_e \dot{I}_p \\ P_{ac} + jQ_{ac} = \dot{V}_p \dot{I}_p^* \end{cases} \quad (A1)$$

where $\dot{V}_e = V_e \angle \theta_e$ is the equivalent voltage of sending-end AC system; $\dot{V}_p = V_p \angle \theta_p$ is the terminal voltage at PCC, $\Delta \dot{V}$ is the line voltage drop; $P_{ac}$ and $Q_{ac}$ are the active power and reactive power of sending-end AC system; $\dot{I}_p$ and $\dot{I}_p^*$ are the branch current injected into PCC by AC system and its reference value.

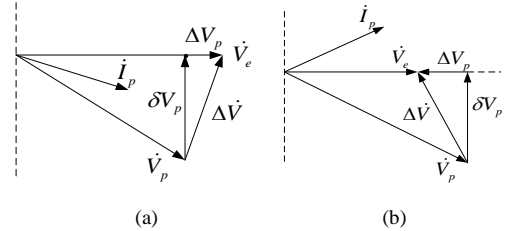

Fig. A1 Phasor diagram of sending-end AC voltage. (a) Normal case. (b) Overvoltage case.

Under normal case where $\dot{V}_p$ is at its rated voltage in Fig. A1(a), the AC system will inject reactive power for HVDC reactive power consumption. When the HVDC system converter is blocked due to some malfunctions such as the continuous commutation failure, the surplus reactive power will be reversely injected into the sending-end, resulting in a transient overvoltage surge at PCC, i.e., $\dot{V}_p$ will be increased to even greater than $\dot{V}_e$, as illustrated in Fig. A1(b), in this case we have:

$$\begin{cases} P_{ac} = -P_{DFIG} \\ Q_{ac} = -Q_{DFIG} - Q_c = -Q_{DFIG} - V_p^2 G_c \end{cases} \quad (A2)$$

$\dot{V}_p$ can further become (A3) by combining (A1) and (A2):

$$\begin{cases} V_p = \sqrt{(V_e + \Delta V_P)^2 + \delta V_P^2} \\ \delta V_P = P_{DFIG} X_e / V_e \\ \Delta V_P = -V_e + (V_p^2 - Q_{DFIG} X_e - V_p^2 G_c X_e)/V_e \end{cases} \quad (A3)$$

$V_e = 1$ is set in (A3). Equation (1) can be easily obtained from (A3).

Considering real voltage profile, $V_p$ can be deduced:

$$V_p = \sqrt{\frac{-b + \sqrt{b^2 - 4ac}}{2a}} \quad (A4)$$

Accordingly, the derivative $\partial V_p / \partial Q_{DFIG}$ and $\partial V_p / \partial P_{DFIG}$ can be obtained as (2) and (3).

The sign of $\partial V_p / \partial Q_{DFIG}$ is analyzed below. Since the SCR of the sending-end AC is defined as:

$$SCR = \frac{S_{sc}}{P_{dN}} = \frac{V_{pN}^2}{X_e P_{dN}} = \frac{1}{X_e} \quad (A5)$$

where $S_{sc}$ is the short circuit capacity of the AC system; $V_{pN}$ and $P_{dN}$ are the base voltage and base power specified in per-unit analysis, respectively; and $P_{dN}$ is the rated HVDC power.

As the reactive compensation capacity $Q_c$ at a HVDC converter station can concisely be expressed as:

$$Q_c = V_{pN}^2 G_c \quad (A6)$$

For HVDC, $Q_c$ usually approximates to $0.4 P_{dN}$ -$0.6 P_{dN}$, thus $G_c \leq 0.6$.

From (2) and (3), it can be found that $\partial V_p / \partial Q_{DFIG} > 0$ holds only if $G_c X_e < 1$, i.e., $SCR > G_c$ or $SCR > 0.6$.

B. Effect of Power Control

Further, $\partial V_p / \partial Q_{DFIG}$ and $\partial V_p / \partial P_{DFIG}$ are compared and analyzed as below.

$$|\frac{\partial V_p}{\partial Q_{DFIG}}| - |\frac{\partial V_p}{\partial P_{DFIG}}| = \frac{(1+\sqrt{b^2-4ac}-2\sqrt{a}P_{DFIG}X_e)X_e}{2\sqrt{a}V_p\sqrt{b^2-4ac}} = \quad (A7)$$
$$k(1+\sqrt{b^2-4ac}-\sqrt{4ac-(b+1)^2})$$

where $k = X_e / (2\sqrt{a}V_p\sqrt{b^2-4ac}) > 0$.

Given the condition $G_c X_e < 1$ and $Q_{DFIG} < 0$, then $b > -1$.

Further, $|\partial V_p / \partial Q_{DFIG}| > |\partial V_p / \partial P_{DFIG}|$ can be deduced. In other words, adding DFIG reactive power absorption has better effect than adding active power on overvoltage suppression.

In summary, when $V_p > V_e$ and $SCR > 0.6$, the following statement can be concluded: adding DFIG reactive power absorption can suppress the transient overvoltage and even has better effect than adjusting active power.

APPENDIX B

The constrain of the reactive power limit of the stator side of DFIG considering the maximum endurable current of RSC is calculated as below.

The voltage, flux and power equations at the stator are listed as (B1)-(B3) by method of decoupling control with d-q axis.

$$\begin{cases} v_{sd} = R_s i_{sd} - \omega_1 \psi_{sq} + \frac{d}{dt}\psi_{sd} \\ v_{sq} = R_s i_{sq} - \omega_1 \psi_{sd} + \frac{d}{dt}\psi_{sq} \end{cases} \quad (B1)$$

$$\begin{cases} \psi_{sd} = L_s i_{sd} + L_m i_{rd} \\ \psi_{sq} = L_s i_{sq} + L_m i_{rq} \end{cases} \quad (B2)$$

$$\begin{cases} P_s = \frac{3}{2}(v_{sd} i_{sd} + v_{sq} i_{sq}) \\ Q_s = \frac{3}{2}(v_{sq} i_{sd} - v_{sd} i_{sq}) \end{cases} \quad (B3)$$

Adopting the stator flux d-axis directional control and neglecting the voltage drop in the stator winding, there is:

$$\begin{cases} \psi_{sd} = \psi_s \\ \psi_{sq} = 0 \end{cases} \quad (B4)$$

$$\begin{cases} v_{sd} = 0 \\ v_{sq} = V_s = \omega_1 \psi_s \end{cases} \quad (B5)$$

Consider the maximum endurable current of RSC as:

$$i_{rd}^2 + i_{rq}^2 \leq (I_r^{max})^2 \quad (B6)$$

From (B1)-(B6), the constrain (4) of the reactive power limit at the stator side of DFIG is obtained.

**Changping Zhou** received the B.Eng. degree in College of Electrical Engineering, Zhejiang University, Hangzhou, China, in 2016, where he is currently working toward the Ph.D. degree. His research interests include power system stability analysis and control of renewable energy.

**Zhen Wang** received the B.Eng., M.Eng., and Ph.D. degrees from Xi'an Jiaotong University, Xi'an, China, Zhejiang University, Hangzhou, China, and Hong Kong Polytechnic University, Hong Kong, China, in 1998, 2001, and 2009, respectively. Currently, he is a full-time professor in the Department of Electrical Engineering, Zhejiang University. He was the recipient of 2014 Endeavour Research Fellowship sponsored by Australia Government and visiting scholar of The University of Western Australia, Perth, Australia, from Feb. 2014 to Aug. 2014. His research interests include power system stability and control, renewable energy integration and VSC-HVDC transmission.

**Ping Ju** received the B.Eng. and M.S. degrees in electrical engineering from Southeast University, Nanjing, China, in 1982, 1985, respectively. He received the Ph.D. degree in electrical engineering from Zhejiang University, Zhejiang, China, in 1988. From 1994 to 1995, he was an Alexander-von-Humboldt Fellow at the University of Dortmund, Germany. He is now a professor of electrical engineering in Hohai University. He is also a Qiushi Chair Professor in Zhejiang University. He has published 6 books and over 300 journal papers. He was awarded the Scientific Funds for Outstanding Young Scientists of China in 2007, and National Science and Technology Progress Award of China in 2017. His research interests include modeling and control of power systems and smart grids with renewable power generation.

**Deqiang Gan** received the Ph.D. degree in electrical engineering from Xi'an Jiaotong University, Xi'an, China, in 1994. Since 2002, he has been with the faculty of Zhejiang University, Hangzhou, China. He visited the University of Hong Kong, Hong Kong, China, in 2004, 2005, and 2006. From 1998 to 2002, he was with the ISO New England, Inc., USA. From 1994 to 1998, he held research positions with Ibaraki University, Mito, Japan, University of Central Florida, Orlando, USA, and Cornell University, Ithaca, USA. He was an editor for European Transactions on Electric Power from 2007 to 2014. His research interests include power system stability and control.